\documentclass[aps,showpacs, nofootinbib]{revtex4}

\newcommand{\be}{\begin{equation}}
\newcommand{\ee}{\end{equation}}
\newcommand{\ben}{\begin{eqnarray}}
\newcommand{\een}{\end{eqnarray}}

\newcommand{\la}{{\lambda}}

\newcommand{\cO}{{\cal O}}

\newcommand{\p}{\partial}
\newcommand{\na}{\nabla}

\newcommand{\tpsi}{\tilde \psi}
\newcommand{\tchi}{\tilde \chi}

\newcommand{\tim}{{\tilde \mu}}
\newcommand{\tom}{{\tilde \omega}}

\newcommand{\Dsl}{{\slash \negthinspace \negthinspace \negthinspace \negthinspace  D}}
\newcommand{\tM}{{\tilde M}}

\newcommand{\tV}{{\tilde V}}
\newcommand{\taa}{{\tilde a}}
\newcommand{\tbb}{{\tilde b}}

\newcommand{\ep}{\epsilon}

\newcommand{\ga}{\gamma}

\pacs{04.50.+h}

\begin{document}

\title{Decay of Massive Dirac Hair on a Brane-World Black Hole}
\author{Gary W.Gibbons}
\affiliation{DAMTP, Centre for Mathematical Sciences,  \protect \\
University of Cambridge\protect \\
Wilberforce Road, Cambridge, CB3 0WA, UK\protect \\
g.w.gibbons@damtp.cam.ac.uk} 

\author{Marek Rogatko and Agnieszka Szyp\l owska}
\affiliation{Institute of Physics \protect \\
Maria Curie-Sklodowska University \protect \\
20-031 Lublin, pl.~Marii Curie-Sklodowskiej 1, Poland \protect \\
rogat@tytan.umcs.lublin.pl \protect \\
rogat@kft.umcs.lublin.pl}

\date{\today}

\begin{abstract}
We investigate the intermediate and late-time behaviour of the massive Dirac spinor
field in the background of static spherically 
symmetric brane-world black hole solutions.
The intermediate asymptotic behaviour of the  massive spinor
field exhibits  a  dependence on the field's 
parameter mass as well as the multiple number of the wave mode.
On the other hand,
the late-time behaviour power law decay has a 
rate which is independent of those factors.
\end{abstract}

\maketitle

\section{Introduction}
Nowadays it is widely believed that  extra dimensions play a
significant 
role in the construction of
a unified theory of the four fundamental forces of nature.
In such models it is often the case that our  Universe can be treated
as a  submanifold to  which the standard model is confined,
embedded in a higher dimensional
spacetime.
If one takes  the volume of the extra dimensions spacetime 
to be sufficiently large, one is able to lower the  fundamental quantum
gravity scale to the electrovac scale of the order of a TeV.
It is thus of interest to construct black hole solutions 
in such brane-world models. The difficulties arising in such  attempts 
stem from the fact that, in general,   brane dynamics 
generates Weyl curvatures which in turn backreact on the brane
dynamics. 
We can look at the problem in question
by projecting the  Einstein equations onto  the brane. 
This approach   was introduced  in Refs.\cite{dad00,cas02}.
It is also of interest to think  of  a four-dimensional
brane-world  black hole solution as a slice that 
intersects a bulk black hole  \cite{kod02,tan03,sea05}. 
In Ref.\cite{gal06}
the possibility was raised of finding a regular  
Randall-Sundrum (RS) brane world on which a static spherically symmetric
black hole, surrounded by realistic matter, is located. 
This was achieved by slicing a fixed five-dimensional bulk black hole 
spacetime. On the other hand, studies of spherically symmetric
brane-world solutions with induced gravity were extended to include
 nonlocal bulk effects \cite{kof02}. 
The scalar as well as the axial gravitational perturbations of what we
shall call ``brane-world black holes''  were studied in Ref.\cite{abd06}.
\par
An  important question for  for black hole physics is 
the  investigation of  how various fields decay
in the spacetime outside   a collapsing body. 
The importance arises from the fact that,   regardless
of the details of the gravitational collapse and features of the 
collapsing body, 
the outcome of this process, i.e.  the resultant  black hole 
is characterized by just a  few parameters such as  mass,
charge and angular momentum. The first researches in this direction were
carried by Price in Ref.\cite{pri72} while
the scalar perturbations on Reissner-Nordstr\o m (RN)
background were considered in \cite{bic72}.
It was found that charged scalar hair
decayed more slowly than  neutral hair  \cite{pir1}-\cite{pir3}, while
the late-time tails in the gravitational collapse of a massive
fields in the background of Schwarzschild solution were 
reported by Burko \cite{bur97}
and in the 
the Reissner-Nordstr\o m solution at intermediate late-time were considered
in Ref.\cite{hod98}. 
The very late-time tails of the massive scalar fields in the
Schwarzschild and nearly extremal Reissner-Nordstr\o m black holes
were obtained  in Refs.\cite{ja}, \cite{ja1}. It was shown that 
the oscillatory
tail of scalar field  decays like  $t^{-5/6}$ at                                late time.
The power-law tails in the evolution of a charged massless scalar
field around the  fixed
background of a dilaton black hole were  
studied in Ref.\cite{mod01a}, while the case of a massive
scalar field was treated in \cite{mod01b}.  
The analytical proof of the intermediate and late-time behaviour of
the  in the case of dilaton gravity with arbitrary coupling constant
was provided in Ref.\cite{rog07}.
On the other hand,
the problem of the late-time behaviour of massive Dirac fields 
were studied respectively
in the spacetime of Schwarzschild, Reissner-Nordstr\o m and 
Kerr-Newman black hole \cite{jin04,jin05,xhe06}.
Ref.\cite{gibrog} was devoted to the analytical studies of the 
intermediate and late-time decay pattern of massive
Dirac hair on a  spherically symmetric dilaton black hole, in 
dilaton gravity theory with arbitrary coupling constant
$\alpha$.
\par
The growth  of interests in unification scheme such as 
superstring/M-theory triggered in turn an  interest in
the decay of hair in the spacetimes of $n$-dimensional black holes.
The {\it no-hair} and uniqueness property  for static holes is by now 
quite well established \cite{unn}. 
The  decay mechanism for  massless scalar  hair  
in the $n$-dimensional Schwarzschild
spacetime was given  in
Ref.\cite{car03}. 
The decay pattern of scalar massive fields in 
the spacetime of $n$-dimensional 
static charged black hole was discussed  in Ref.\cite{mod05}. 
It was shown  that the intermediate
asymptotic behaviour of the hair in question was of the 
form $t^{- (l + n/2 -1/2)}$. Numerical experiment
for $n = 5$ and $n = 6$ confirmed these  results.
In Ref.\cite{cho07a} the authors obtained 
fermion quasi-normal modes for massless Dirac fermion in the 
background of higher dimensional Schwarzschild black hole.
\par
As far as the brane-world  black holes are  concerned,
Ref.\cite{rog07br} was devoted to studies of the intermediate and 
late-time behaviour of the massive
scalar field in the background of a static spherically symmetric 
brane-world  black hole. Among other things,  it was shown that the
late-time power law decay rate is proportional to $t^{-5/6}$.
The massless fermion excitations on a tensional 3-brane embedded in 
six-dimensional spacetime were studied in
\cite{cho07b}.\\
The main aim of our paper will be to clarify what kind of 
mass-induced behaviour  plays  the dominant role
in the asymptotic late-time tails as a result of  decaying 
 massive Dirac spinor hair in the background of brane-world black hole.
\par
The paper is organized as follows.
In Sec.II we gave the analytic arguments concerning
the decay of massive Dirac hair in the background of the considered black hole. 
Sec.III will be devoted to a summary and discussion.

\section{The Decay of Dirac Hair in the Background of Black Hole Brane Solution}
\subsection{Spinor fields}
We shall begin our analysis by recalling the general properties of
massive Dirac equation in an $n$-dimensional
spherically symmetric background \cite{gibrog}. Namely, we shall study the massive Dirac Eq. given by the relation
\be
\bigg(\ga^{\mu} \na_{\mu} \psi - m \bigg) \psi = 0,
\ee
where $\na_{\mu}$ is the covariant derivative 
$\na_{\mu} = \p_{\mu} + {1\over 4} \omega_{\mu}^{ab} \ga_{a} \ga_{b}$, $\mu$
and $a$ are tangent and spacetime indices. There are related by $e_{\mu}^{a}$, a basis 
of orthonormal one-forms. The quantity $\omega_{\mu}^{ab} \equiv \omega^{ab}$ are the associated connection
one-forms satisfying
$de^{a} + \omega_{b}{}{}^{a} \wedge e^{b} = 0$. On the other hand, $\ga^{\mu}$ are Dirac matrices 
fulfilling relation $\{ \ga^{a}, \ga^{b} \} =  2 \eta^{ab}$.
\par
If a  metric  takes the product 
 form :
\be
g_{\mu \nu} d x ^\mu d x ^\nu = 
 g_{ab} (x) dx^a dx ^b + g_{mn}(y) dy^m dy ^n\,,  
\ee
 then Dirac operator $\Dsl$ satisfies a  direct sum decomposition
\be
\Dsl = \Dsl_x + \Dsl_y.
\ee
If one defines a Weyl conformally rescaled metric by 
 $g_{\mu \nu} = \Omega ^2 {\tilde g} _{\mu \nu}$
one finds that    
\be
\Dsl \psi =\Omega ^{- {1 \over 2} (n+1)} { \tilde {\Dsl} } {\tilde \psi},  
\qquad \psi = \Omega ^ {- {1 \over 2}  (n-1) }\tilde \psi.
\ee
Because a  spherically symmetric line element 
is necessarily conformally flat,  
a static metric, spherically symmetric metric of the form 
\be
ds^2 = -A^2 dt^2 + B^2 dr ^2 + C^2 d \Sigma ^2 _{n-2} 
\ee
where $A=A(r), B=B(r), C=C(r)$ are functions only of the radial variable
$r$, and the {\it transverse} metric $d \Sigma ^2 _{n-2} $  
depends neither on $t$ nor on  $r$ is conformal to an ultrastatic 
metric, one factor of which is conformally flat. This allows us to
solve the Dirac equation by a succession of conformal transformations
and direct sum decompositions.  
The assumption that $\Psi$ is a spinor eigenfunction on the $(n-2)$-dimensional {\it transverse} manifold 
$\Sigma$, leads to the equation:
\be
\Dsl_\Sigma \Psi = \lambda \Psi.
\ee
In case of $(n - 2)$-dimensional sphere the eigenvalues for spinor $\Psi$,
where found in Ref.\cite{cam96}. They imply
\be
\la^2 = \bigg( l + { n - 2 \over 2} \bigg)^2,
\ee
where $l = 0, 1, \dots$ \\
Having in mind the properties given above, one may suppose that
\be
\Dsl \psi = m \psi, 
\label{mm}
\ee 
and take  the form of the spinor $\psi$ to be :
\be
\psi = {1 \over A^{1 \over 2} } { 1 \over C^{(n-2) \over 2 }} \chi \otimes \Psi.
\ee
If one carries out the explicit calculations it turns  out that :
\be
(\gamma ^0 \partial _t + \gamma ^1 \partial _x) \chi = A (m- {\lambda \over C} ) \chi,  
\ee
where we have denoted by 
\be
dy = {B \over A} dr,
\ee
the radial optical distance (i.e., the Regge-Wheeler radial coordinate).
On the other hand, the gamma matrices $\gamma ^0, \gamma ^1$ 
satisfy the Clifford algebra in two spacetime dimensions.
One should remark that
having in mind a Yang-Mills  gauge field $A_{\mu}$,
an identical result can be provided 
on the {\it transverse} manifold $\Sigma$. Namely, we have  
\be
\Dsl_{\Sigma, A_{\mu}}  \Psi = \lambda \Psi,
\ee
where $ \Dsl_{\Sigma, A_{\mu}} $ is 
the Dirac operator twisted by the the connection $A_{\mu}$.

Finally, if we take into account that $\psi$ has the form as $\psi
\propto e^{-i\omega t}$ one obtains a 
second order equation for $\chi$, that is 
\be
{d^2 \chi \over d y^2 }+ \omega ^2 \chi  = 
A^2 (m-{\lambda \over C} )^2 \chi. 
\label{second}
\ee

\subsection{Dadhich-Maartens-Papadopoulous-Rezania (DMPR) brane-world black hole solution}
We treat first the case of
the static spherically symmetric black hole localized on a three-brane
in five-dimensional gravity in Randall-Sundrum model \cite{ran99}.
Having in mind the effective field equations on the brane one gets the
following brane-world  black hole metric
\cite{dad00}:
\be
ds^2 = - \bigg( 1 - {2M \over M_{p}^2~r} + {q^2 \over \tM_{p}^2~r^2} \bigg) dt^2 +
{dr^2 \over \bigg( 1 - {2M \over M_{p}^2~r} + {q^2 \over \tM_{p}^2~r^2} \bigg)} + r^2~d\Omega^2,
\label{dila}
\ee
where $q$ is a dimensionless tidal parameter arising from the projection onto the brane of the gravitational
field in the bulk, $\tM_{p}$ is a fundamental five-dimensional Planck mass while $M_{p}$ is the effective Planck mass
in the brane world. Typically, one has  $\tM_{p} \ll M_{p}$. 
 In what follows we shall concentrate on the negative tidal charge which is claimed \cite{dad00}
to be the more natural case. Thus, the roots of $g_{00} = 0$ are respectively $r_{+}$ and $r_{-}$. Namely, they imply
\be
r_{\pm} = {M \over \tM_{p}^2 }\bigg( 
1 \pm \sqrt{1 - {q M_{p}^4 \over M^2~\tM_{p}^2}} 
\bigg).
\ee
Expressing the negative charge as $Q$, for simplicity, we can rewrite the roots as follows:
\be
r_{\pm} = M \bigg( 1 \pm \sqrt{1 + {Q \over M^2}} \bigg).
\ee
\par
Our main aim will be to
analyze the time evolution of a massive Dirac spinor field in the background of
brane-world  black hole by means of the spectral decomposition method.
In Refs.\cite{hod98},\cite{lea86} 
it was argued that the asymptotic massive tail is due to 
 the existence of a 
branch cut placed along the interval $-m \le \omega \le m$.
Thus,
an oscillatory inverse power-law behaviour of the massive spinor field arises
from the integral of Green function $\tilde G(y, y';\omega)$ around
the  branch cut.
Consider, next, the time evolution of the massive
 Dirac spinor field provide by the relation
\be
\chi(y, t) = \int dy' \bigg[ G(y, y';t) \chi_{t}(y', 0) +
G_{t}(y, y';t) \chi(y', 0) \bigg],
\ee
for $t > 0$, where   the Green's function  $ G(y, y';t)$ implies
\be
\bigg[ {\p^2 \over \p t^2} - {\p^2 \over \p y^2 } + V \bigg]
G(y, y';t)
= \delta(t) \delta(y - y').
\label{green}
\ee
By means of the Fourier transform \cite{lea86}
$\tilde  
G(y, y';\omega) = \int_{0^{-}}^{\infty} dt~ G(y, y';t) e^{i \omega t}$, 
Eq.(\ref{green})
can be reduced to an ordinary differential equation.
The Fourier  transform is well defined for $Im~ \omega \ge 0$, while the 
corresponding inverse transform yields
\be
G(y, y';t) = {1 \over 2 \pi} \int_{- \infty + i \ep}^{\infty + i \ep}
d \omega~
\tilde G(y, y';\omega) e^{- i \omega t},
\ee
for some positive number $\ep$.
By virtue of the above
the Fourier  component of the Green's function $\tilde  G(y, y';\omega)$
can be rewritten in terms of two linearly independent solutions for
homogeneous equation.
Namely, it reduces to 
\be
\bigg(
{d^2 \over dy^2} + \omega^2 - \tV \bigg) \chi_{i} = 0, \qquad i = 1, 2,
\label{wav}
\ee
where $\tV = A^{2} \bigg( m - {\la \over C} \bigg)^2$.\\

The boundary conditions for $\chi_{i}$ are described by purely ingoing waves
crossing the outer horizon $H_{+}$ of the 
static black hole
$\chi_{1} \simeq e^{- i \omega y}$ as $y \rightarrow  - \infty$, while 
$\chi_{2}$ should be damped exponentially at $i_{+}$. Thus,
$\chi_{2} \simeq e^{- \sqrt{m^2 - \omega^2}y}$ at $y \rightarrow \infty$.
\par
Suppose now that the observer and the initial data are situated far away from the considered brane
black hole. Let us rewrite Eq.(\ref{wav}) in the more convenient form using
the change of variables
\be
\chi_{i} = {\xi \over \bigg( 1 - {r_{+} \over r} \bigg)^{1/2}
\bigg( 1 - {r_{-} \over r} \bigg)^{1/2}},
\ee
where $i = 1,2$. Then,
we expand Eq.(\ref{wav}) as a power  series of $ r_{\pm}/r$ neglecting terms of order
$\cO ((\omega /r)^2)$ and higher. We obtain the the following:
\ben \label{wavv}
{d^2 \over dr^2} \xi &+& \bigg[
\omega^2 - m^2 + {2 \omega^2 ( r_{+} +  r_{-}) 
- m^2 ( r_{+} + r_{-})
+ 2 \la m (1 + r_{+})
\over r} \\ \nonumber
&-& {\la^2 - 2 \la m r_{-} + m^2 r_{+} r_{-}
\over r^2}
\bigg] \xi = 0.
\een
The solution of equation (\ref{wavv}) may be obtained in terms of 
Whittaker  functions. 
Two basic solutions are needed to construct the Green function, 
with the condition that
$\mid \omega \mid \ge m$, i.e., $\tchi_{1} = M_{\delta, \tim}(2 \tom r)$ and $\tchi_{2} = W_{\delta, \tim}(2 \tom r)$.
The parameters of them imply
\ben
\tim &=& \sqrt{ 1/4 + \la^2 - 2 \la m r_{-} + m^2 r_{+} r_{-} }, \\ \nonumber
\delta &=&  {\omega^2 ( r_{+} + r_{-}) + \la m (1 + r_{+}) - {m^2 \over 2}( r_{+} +  r_{-})
\over \tom},\\ \nonumber
\tom^2 &=& m^2 - \omega^2.
\een
Having all this in mind, we reach to the following form of the spectral Green function:
\ben
G_{c}(x,y;t) &=& {1 \over 2 \pi} \int_{-m}^{m}dw
\bigg[ {\tchi_{1}(x, \tom e^{\pi i})~\tchi_{2}(y,\tom e^{\pi i}) \over W(\tom e^{\pi i})}
- {\tchi_{1}(x, \tom )~\tchi_{2}(y,\tom ) \over W(\tom )} 
\bigg] ~e^{-i w t} \\ \nonumber
&=& {1 \over 2 \pi} \int_{-m}^{m} dw f(\tom)~e^{-i w t},
\een 
where $W(\tom)$ stands for Wronskian.\\
Let us analyze first,
the intermediate asymptotic behaviour of the massive spinor field with the range of parameters
$M \ll  r \ll t \ll M/(m M)^2$.
The intermediate asymptotic contribution to the Green function integral gives the frequency equal to 
$\tom = {\cO (\sqrt{m/t})}$, which in turns implies that $\delta \ll 1$. 
Using the fact 
that $\delta$ 
results from the $1/r$ term in the massive spinor field equation of
 motion, it illustrates
the effect of backscattering off the spacetime curvature.
In the case under consideration
the backscattering is negligible. Thus, we find the following:
\be
f(\tom) = {2^{2 \tim -1} \Gamma(-2\tim)~\Gamma({1 \over 2} + \tim) \over
\tim \Gamma(2 \tim)~\Gamma({1 \over 2} - \tim)} \bigg[
1 + e^{(2 \tim + 1) \pi i} \bigg]
(r r')^{{1 \over 2} + \tim} \tom^{2 \tim},
\ee
where one applied the fact that $\tom r \ll 1$.
We also have in mind that 
$f(\tom)$
can be approximated using  the fact that $M(a, b, z) = 1$ as $z$ tends to zero.
Consequently, the resulting spectral Green function reduces to the form as
\be
G_{c}(r, r';t) = {2^{3 \tim - {3\over2}} \over \tim \sqrt{\pi}}
{\Gamma(-2\tim)~\Gamma({1 \over 2} + \tim) \Gamma(\tim +1 ) \over
\tim \Gamma(2 \tim)~\Gamma({1 \over 2} - \tim)}
\bigg( 1 + e^{(2 \tim + 1) \pi i} \bigg)~(r r')^{{1 \over 2} + \tim} 
~\bigg( {m\over t} \bigg)^{{1 \over 2} + \tim}~J_{{1 \over 2} + \tim}(mt).
\ee  
Taking into account the limit when $t \gg 1/m$ we conclude that the spectral Green function yields
\be
G_{c}(r, r';t) = {2^{3 \tim - 1} \over \tim \sqrt{\pi}}
{\Gamma(-2\tim)~\Gamma({1 \over 2} + \tim) \Gamma(\tim +1 ) \over
\tim \Gamma(2 \tim)~\Gamma({1 \over 2} - \tim)}
\bigg( 1 + e^{(2 \tim + 1) \pi i} \bigg)~(r r')^{{1 \over 2} + \tim} 
~m^{\tim}~ t^{- 1 - \tim}~\cos(mt - {\pi \over 2}(\tim + 1)).
\label{gfim}
\ee  
We remark that Eq.(\ref{gfim}) exhibits 
an  oscillatory inverse power-law behaviour. In our case the intermediate
times of the power-law tail depends only on
 $\tim$ which in turn is a function of the multiple number
of the wave modes.
\par
The other pattern of decay of massive spinor Dirac hair is expected 
when  $\delta \gg 1$, for the late-time
behaviour. Namely, when the backscattering off the curvature is taken
 into account.
Under the assumption that  $\delta \gg 1$, 
$f(\tom)$ may be rewritten in the form as
\ben \label{fer}
f(\tom) &=& {\Gamma(1 + 2\tim)~\Gamma(1 - 2\tim) \over 2 \tim}~(r r')^{1 \over 2}
\bigg[ J_{2 \tim} (\sqrt{8 \delta \tom r})~J_{- 2 \tim} (\sqrt{8 \delta \tom r'})
- I_{2 \tim} (\sqrt{8 \delta \tom r})~I_{- 2 \tim} (\sqrt{8 \delta \tom r'}) \bigg] \\ \nonumber
&+&
{(\Gamma(1 + 2\tim))^2~\Gamma(- 2\tim)~\Gamma( {1 \over 2} + \tim - \delta)
 \over 2 \tim ~\Gamma(2 \tim)~\Gamma({1 \over 2} - \tim - \delta) }~(r r')^{1 \over 2}
~\kappa^{- 2 \tim}
\bigg[
J_{2 \tim} (\sqrt{8 \delta \tom r})~J_{2 \tim} (\sqrt{8 \delta \tom r'})
\\ \nonumber
&+& e^{(2 \tim + 1)}
I_{2 \tim} (\sqrt{8 \delta \tom r})~I_{2 \tim} (\sqrt{8 \delta \tom r'}) 
\bigg],
\een
where we used the limit $M_{\delta, \tim}(2 \tom r) \approx
\Gamma (1 + 2 \tim) (2 \tom r)^{1 \over 2}~\delta^{- \tim}~J_{\tim}(\sqrt{8 \delta \tom r})$.
One should notice that
the first part of Eq.(\ref{fer}), the late time tail, is proportional to $t^{-1}$.
It occurs that we shall concentrate on
the second term of the right-hand side of Eq.(\ref{fer}). For the case under consideration
it can be brought to the form:
\be
G_{c~(2)}(r, r';t) = {M \over 2 \pi} \int_{-m}^{m}~dw~e^{i (2 \pi \delta - wt)}~e^{i \varphi},
\ee
where the phase $\varphi$ is defined by the relation
\be
e^{i \varphi} = { 1 + (-1)^{2 \tim} e^{- 2 \pi i \delta} \over
 1 + (-1)^{2 \tim} e^{2 \pi i \delta}},
\ee
while $M$ is given by:
\be
M = {(\Gamma(1 + 2\tim))^2~\Gamma(- 2\tim) \over 2 \tim ~\Gamma(2 \tim) }~(r r')^{1 \over 2}
\bigg[
J_{2 \tim} (\sqrt{8 \delta \tom r})~J_{2 \tim} (\sqrt{8 \delta \tom r'})
+ I_{2 \tim} (\sqrt{8 \delta \tom r})~I_{2 \tim} (\sqrt{8 \delta \tom r'}) 
\bigg].
\ee
The saddle-point integration allows one to find accurately
the asymptotic behaviour. This method is applicable in our case because of
the fact that
at very late time both terms $e^{i w t}$ and $e^{2 \pi \delta}$ are rapidly
oscillating,
which in turns means 
that the spinor waves are mixed states consisting of the states 
with multipole phases backscattered by spacetime curvature. Most of them cancel
with each others which have the inverse phase. 
The saddle-point is found to exist at the following value:
\be
a_{0} = \bigg[ {\pi~(\omega^2~ (r_{+} + r_{-}) +
\la m (1 + r_{-})  - {m^2 \over 2} (r_{+} + r_{-}) 
\over  \sqrt{2} m} \bigg]^{1 \over 3},
\ee
Then, the resultant
form of the spectral Green function yields
\be            
G_{c}(r, r';t) = { 2 \sqrt{2} \over \sqrt{3}}~m^{2/3}~ (\pi)^{5 \over 6}
\bigg[
2 m^2 (r_{+} + r_{-}) + 2 \la m (1 + r_{+}) - m^2 (r_{+} + r_{-}) 
\bigg]^{1 \over 3}
(mt)^{-{ 5 \over 6}}~\sin(mt)~\tchi(r, m)~\tchi(r', m).
\ee
The above form of the spectral Green function concludes our investigations of the late-time
behaviour of massive Dirac hair in the background of DMPR brane-world
 black hole.
The form of it envisages the fact that the late-time behaviour of the fields in question is independent on
the field parameter mass as well as the number of the wave mode. The late-time pattern of decay is proportional to
$- 5/6$.

\subsection{Casadio-Fabbri-Mazzacurati (CFM) brane black hole solution}
Our next task will be to consider a general class of spherically symmetric static solution to five-dimensional
equations of motion by considering the general form of the line element provide by the metric
\be
ds^2 = - A(r) dt^2 + {1 \over B(r)}dr^2 + r^2 d\Omega^2.
\ee
Casodio {\it et al.} \cite{cas02} obtained two types of analytic solutions by fixing either $A(r)$ or $B(r)$.
The solution will be given in terms of the ADM mass $M$ and the parametrized post-Newtonian (PPN) parameter $\beta$
which affects the perihelion shift and the Nordtvedt effect \cite{wil93}. 
The momentum constraints are identically satisfied by the metric coefficients and the {\it Hamiltonian}
constraints can be written out \cite{cas02}. Setting $A(r) = \bigg( 1 - {2 M \over r} \bigg)$ the resulting
metric yields
\be
ds^2 = - \bigg( 1 - {2 M \over r} \bigg) dt^2 +
{\bigg( 1 - {3 M \over 2 r} \bigg) \over
\bigg(1 - {2 M \over r} \bigg)~\bigg( 1 - {\gamma M \over 2 r} \bigg)} dr^2
+ r^2 d\Omega^2,
\ee
where $\gamma = 4 \beta -1$.
A convenient form of the equation of motion for a 
 massive Dirac field can be obtained by the transformation
:
\be
\chi_{i} = {\bigg( 1 - {3M \over 2r} \bigg)^{1 \over 4}
\over \bigg( 1 - {2M \over r} \bigg)^{1/2}
\bigg( 1 - {\gamma M \over 2r} \bigg)^{1/4}} \xi,
\ee
where $i = 1,2$. As in the preceding section,
let us expand Eq.(\ref{green}) as a power  series of $ M/r$ neglecting terms of order
$\cO ((\omega/r)^2)$ and higher. It then follows directly that one has
\be
{d^2 \over dr^2} \xi + \bigg[
\omega^2 - m^2 +
{\omega^2~\taa - m^2~\tbb + 2 \la m  \over r}
 - {\la^2 + \la~ M~ m~ (3 - \gamma) - {3 \over 4} M^2~\gamma~ m^2 \over r^2}
\bigg] \xi = 0,
\ee
where $\taa = {M \over 2}~(5 + \ga)$ and $\tbb = {m \over 2}~(\gamma - 3)$.\\

Thus, the two basic solutions which are needed to construct the Green function, with the condition that
$\mid \omega \mid \ge m$ are given by $\tchi_{1} = M_{\delta, \tim}(2 \tom r)$ and $\tchi_{2} = W_{\delta, \tim}(2 \tom r)$,
with the following parameters:
\be
\tim = \sqrt{ 1/4 + \la^2 + \la~ M~ m~ (3 - \gamma) - {3 \over 4} M^2~\gamma~ m^2 }, 
\qquad    \delta = {\omega^2~\taa - m^2 ~\tbb + 2 \la m \over 2 \tom } \qquad        
\tom^2 = m^2 - \omega^2.
\label{casea}
\ee
The preceding section arguments can be repeated. The conclusion is that 
the spectral Green function of the intermediate late-time behaviour
of massive Dirac spinor fields with the  new parameters of
the  Whittaker functions given by the relation (\ref{casea}).
Consequently, 
the next step will be to calculate the late-time behaviour of the considered field. It can be verified that 
the
stationarity of the  integral will be achieved for the parameter
\be
a_{0} = \bigg[ 
{ \pi~( \omega^2~\taa - m^2~\tbb + 2 \la m ) \over 2 \sqrt{2} m}
 \bigg]^{1 \over 3}.
\ee
By virtue of saddle point method, on evaluating the adequate expressions, we find that
the spectral Green function provides the following:
\be            
G_{c}(r ,r';t) =  {2 \sqrt{2} \over \sqrt{3}}~m^{2/3} (\pi)^{5 \over 6}
\bigg[ 4M m^2 + 2 \la m \bigg]^{1 \over 3}~
(mt)^{-{ 5 \over 6}}~\sin(mt)~\tpsi(r, m)~\tpsi(r', m),
\ee
One can observe that the dominant role in the late-time 
behaviour is  played by  the term
proportional to $- 5/6$.\\
On the other hand, let us consider that $B(r) = \bigg( 1 - {2 \ga M \over r} \bigg)$
for the other model of brane-world  black hole. It implies the following line element:
\be
ds^2 = {1 \over \ga} \bigg( \ga -1 + \sqrt{1 - {2 \ga M \over r}} \bigg)^2 dt^2 +
{dr^2 \over \bigg( 1 - {2 \ga M \over r} \bigg)} + r^2 d \Omega^2.
\ee
Next, let us change  coordinates  as follows:
\be
\chi_{i} = {\ga^{1 \over 2} \xi \over
\bigg( \ga - 1 + \sqrt{1 - {2 \ga M \over r}} \bigg)^{1 \over 2}
\bigg(1 - {2 \ga M \over r} \bigg)^{1 \over 2}},
\ee
where $i = 1,2$,
Then, expand Eq.(\ref{green}) as a power  series of $ M/r$ neglecting terms of order
$\cO ((\omega/r)^2)$ and higher. It yields
\be
{d^2 \over dr^2} \xi + \bigg[ \omega^2 ~\ga^2~ \rho^2 - m^2 +
{4 \ga M ( \omega^2~ (1 + \rho^2) - m^2) + 2 \la m \over r}
 -{\la^2  - 8 m M \la \gamma - 4 M^2 m^2 \gamma^2 \over r^2}
\bigg] \xi = 0,
\label{whita}
\ee
where $\rho^2 = (\ga - 1)^2 + 3$.\\
Eq.(\ref{whita}) can be brought to the form of  Whittaker's equation. Two basic solutions are needed to construct the Green function.
The additional requirement
that
$\mid \omega \mid \ge m$, implies that they are of the form 
$\tchi_{1} = M_{\delta, \tim}(2 \tom r)$ and $\tchi_{2} = W_{\delta, \tim}(2 \tom r)$.
The parameters of the Whittaker functions are given by
\be
\tim = \sqrt{ 1/4 + \la^2 - 8 m M \la \gamma - 4 M^2 m^2 \gamma^2 }, 
\qquad    \delta = {4 \ga M [\omega^2 (1 + \rho^2) - m^2] + 2 \la m
\over 2 \tom}
 \qquad        
\tom^2 = m^2 - \omega^2~\ga^2~\rho^2.
\ee
On the other hand, from the considerations presented in the preceding case,
the stationarity of $2 \pi \delta - \omega t$ can be obtained for the parameter equal to 
\be
a_{0} = \bigg[ { \pi~(4 \ga M ( \omega^2 (1 + \rho^2)  - m^2 ) + 2 \la m)
 \over 2 \sqrt{2} m} \bigg]^{1 \over 3}.
\ee
Summing it all up, one obtains the asymptotic late-time spectral Green function in the form
\be
G_{c}(r , r';t) =  {2\sqrt{2} \over \sqrt{3} \ga^2 [(\ga - 1)^2 + 3 ]}~m^{2/3}~ (\pi)^{5 \over 6} 
\bigg[ 4 \ga M m^2 ((\ga - 1)^2 + 3) + 2 \la m
\bigg]^{1 \over 3}~
(mt)^{-{ 5 \over 6}}~\sin(mt)~\tchi(r, m)~\tchi(r', m),
\ee
As in the previous cases the dominant role in the asymptotic late-time decay of massive
Dirac hair in the spacetime of CFM brane black hole plays the oscillatory tail with the decay rate proportional to
$t^{-5/6}$.

\section{Conclusions}
In our paper we have considered the problem of the asymptotic tail behaviour of massive Dirac 
hair in the spacetime of various brane-world  black hole solutions. 
Our main aim was to reveal what type
of mass-induced behaviours play the main role in the 
asymptotic intermediate and
late-time decay pattern of black hole hair in question. In our considerations
we took into account two brane-world  black hole solutions given in 
Refs.\cite{dad00,cas02}.
It was shown that in the case of intermediate asymptotic behaviour one gets the oscillatory
power-law dependence
which varied with the multiple number of the wave mode $l$ as well as with the mass of the Dirac fields.
As in the case of {\it ordinary} Einstein static spherically symmetric black hole spacetimes
this pattern of decay is not the final one. At very late-times the resonance backscattering off
the spacetime curvature emerges, which in turn is independent on
angular momentum parameter and the field parameter $m$.
The late-time asymptotic pattern of the decay is of the form $t^{- 5/6}$. 
It should be interesting to find a general proof of this pattern of decay for massive
Dirac fields in the spacetime of static spherically symmetric black object.
The investigations in this direction is in progress and will be published elsewhere.

\begin{acknowledgments}
MR is grateful for hospitality of DAMTP, Center for Mathematical Sciences, Cambridge,
were the part of the research was started. This work was partially financed by the Polish
budget funds in 2007 year as the research project.

\end{acknowledgments}




\end{document}